\documentclass[12pt,a4paper]{iopart}

\newcommand{\eq}[1]{\begin{equation}#1\end{equation}}
\newcommand{\dd}{\mathrm{d}}
\newcommand{\ee}{\mathrm{e}}

\expandafter\let\csname equation*\endcsname\relax
\expandafter\let\csname endequation*\endcsname\relax 
\usepackage{amsmath}
\usepackage{iopams}
\usepackage{graphics}
\usepackage{graphicx}
\usepackage{psfrag}
\usepackage{color}
\usepackage{colordvi}

\begin{document}

\title{Properties of the entanglement Hamiltonian for finite free-fermion chains}
\author{ Viktor Eisler$^1$ and Ingo Peschel$^2$}
\address{
$^1$Institut f\"ur Theoretische Physik, Technische Universit\"at Graz, Petersgasse 16,
A-8010 Graz, Austria\\
$^2$Fachbereich Physik, Freie Universit\"at Berlin, Arnimallee 14, D-14195 Berlin, Germany
}

\begin{abstract}
  We study the entanglement Hamiltonian for fermionic hopping models on rings and open chains
  and determine single-particle spectra, eigenfunctions and the form in real space. For the
  chain, we find a commuting operator as for the ring and compare with its properties
  in both cases. In particular, a scaling relation between the eigenvalues is found for large
  systems. We also show how the commutation property carries over to the critical transverse Ising
  model.
 \end{abstract}

\footnotetext{
{\em Dedicated to the memory of Vladimir Rittenberg.}}

\maketitle

\section{Introduction}

If one studies the entanglement properties of a quantum state \cite{CCD09,Laflorencie16},
the central quantity is the reduced density matrix of one of the two subsystems. It can always
be written in the form $\rho=\exp(-\mathcal{H})/Z$, and the operator $\mathcal{H}$ is now
commonly called the entanglement Hamiltonian \cite{Li/Haldane08}. Its structure has been the
subject of various recent studies, both in the continuum
\cite{Casini/Huerta/Myers11,Wong_etal13,Wen_etal16, Cardy/Tonni16,
Pretko17,Arias_etal17_1,Arias_etal17_2,Klich/Vaman/Wong17_1,Klich/Vaman/Wong17_2,Tonni/Laguna/Sierra17}
and on the lattice
\cite{Peschel/Eisler09,Nienhuis/Campostrini/Calabrese09,Kim_etal16,Eisler/Peschel17,Tonni/Laguna/Sierra17,
Dalmonte/Vermersch/Zoller17,Parisen/Assaad18}.

For a subsystem of length $\ell$ in a critical quantum chain, conformal field theory gives for
various cases a result of the form 
\cite{Cardy/Tonni16}
\begin{equation}
\mathcal{H} = 2\pi \ell \int_0^{\ell} dx \; \beta(x)\;T_{00}(x) \, ,
\label{conf_ham}
\end{equation}
where $T_{00}$ is the energy density in the physical Hamiltonian \cite{Casini/Huerta/Myers11,Wong_etal13,
Wen_etal16, Cardy/Tonni16} and $\beta(x)$ is a weight factor arising from the conformal mapping
which relates the path integral in the actual geometry to that for a strip. It has
been viewed as a local inverse temperature and used to obtain the entanglement entropy from thermodynamic
relations
\cite{Wong_etal13,Pretko17,Arias_etal17_1,Arias_etal17_2}.
For an interval in an infinite chain, it is a parabola
$\beta(x)=x/\ell \,(1-x/\ell)$ vanishing linearly at the ends of the interval.

On the lattice, the situation is somewhat different and intriguing. On the one hand, $\mathcal{H}$ for
an interval in an infinite hopping chain does \emph{not} have exactly the conformal form
\cite{Eisler/Peschel17}, but on the other hand there exists a commuting operator $\mathcal{T}$ which
is rather close to $\mathcal{H}$ and \emph{does} have this form.

In the present communication, we study $\mathcal{H}$ for \emph{finite} free-fermion systems which are
either rings or open chains. In a recent investigation, this was done already for an inhomogeneous model,
the so-called rainbow chain, for which also conformal results are available \cite{Tonni/Laguna/Sierra17}.
Here, however, we are interested in the simpler homogeneous case, where a commuting operator exists
and can serve as a point of reference. For the ring (as for the infinite system), this property has been known
for quite a while, albeit in a different context \cite{Grünbaum81,Xu/Chamzas84}.
For a (simply divided) open chain, the result is new and will be derived in Section 3. As for the ring, the
commuting operator has exactly the conformal form.

In dealing with a finite discrete system, one encounters a particular feature. If the particle number
is too small or too large, some states in the subsystem are definitely empty or definitely filled. For
calculations of the entanglement entropy as in \cite{Fagotti/Calabrese11}, these play no role, but in
the entanglement Hamiltonian they lead to infinities, and $\mathcal{H}$ is not well defined.
Excluding such cases, we determine the properties of $\mathcal{H}$ from high-precision
diagonalizations of the correlation matrix as in \cite{Eisler/Peschel17} and compare them to those of
the infinite systems and to those of the commuting operator $\mathcal{T}$. Here two features are to be
mentioned. The single-particle eigenfunctions of the finite and the infinite case are almost identical,
while the eigenvalues differ clearly. Also the nearest-neighbour hopping in $\mathcal{H}$ decreases
significantly near the end of an open chain, in contrast to the conformal result. We also discuss
formulae for the spectra for large $\ell$ and the connection between the eigenvalues of $\mathcal{H}$
and of $\mathcal{T}$ in analogy to the infinite case. Finally, it is shown that all the features of
the hopping model can also be found in the critical transverse-field Ising (TI) model by properly
relating both following \cite{Igloi/Juhasz08}. In particular, there is an operator commuting with
$\mathcal{H}$ also in this case.
 
In the following Section 2 we formulate the problem and give the basic expressions.
In Section 3, we present the determination of the
commuting operator for the open chain and compare in Section 4 the properties of $\mathcal{H}$
and $\mathcal{T}$. In Section 5 we discuss some asymptotic properties, while Section 6 contains a
summary. Finally, in the appendix, we outline the steps necessary to map the TI model to the
hopping model.

\section{Setting}

We consider the ground state of a system of spinless fermions hopping on a finite chain of $L$ sites
$n=1,2,\dots,L$ with Hamiltonian
\begin{equation}
\mathcal{\hat H} = -\frac{1}{2} \sum_{n=1}^{L'} \,t\,(c^{\dag}_n c_{n+1} + 
c^{\dag}_{n+1} c_{n})+\sum_{n=1}^L\,d \, c^{\dag}_n c_n = \sum_{m,n=1}^L \hat H_{m,n}
c^{\dag}_m c_n \, .
\label{hamhop}
\end{equation}
We take the hopping $t=1$ and use the site energy $d$ to adjust the ground-state filling.
For a ring with periodic boundary conditions, $L'=L$ in the first sum, 
while for a chain with open ends, $L'=L-1$.

To diagonalize $\mathcal{\hat H}$, one has to find the eigenfunctions $\Phi_q$ and
eigenvalues $\omega_q$ of the matrix $\hat H$. For a ring this gives
\begin{equation}
  \Phi_q(n)= \sqrt \frac{1}{L} \exp (iqn)  \, , \quad \quad \quad q=\frac{2\pi k}{L},\quad
              k=0, \pm 1, \pm 2,\dots,+L/2 \, ,
\label{eigen_ring}
\end{equation}
while for the open chain
\begin{equation}
  \Phi_q(n)= \sqrt \frac{2}{L+1} \sin (qn)  \, , \quad \quad \quad q=\frac{\pi k}{L+1},\quad k=1,2,\dots,L \, .
\label{eigen_chain}
\end{equation}
The eigenvalues are given in both cases by $\omega_q=-\cos q + d$.
The entanglement properties are then determined by the correlation matrix 
$C_{m,n}$ in the ground state
\begin{equation}
  C_{m,n} = \langle c^{\dag}_m c_{n} \rangle = \sum_{|q|<q_F} \Phi^*_q(m)\Phi_q(n) \, , 
\label{corr_gen}
\end{equation}
where the sum is over the occupied states. This gives for the ring 
\begin{equation}
  C_{m,n} = \frac{\alpha}{\pi} \frac{\sin q_F(m-n)}{\sin\alpha(m-n)}
         \,,\quad \quad \quad \alpha= \pi/L \, ,
\label{corr_ring}
\end{equation}
where $q_F= 2\alpha(K+1/2)$ if the $2K+1$ momenta with $k=0,\pm 1,\pm 2,\dots,\pm K$ are occupied.
For the chain, one finds \cite{Fagotti/Calabrese11}
\begin{equation}
\hspace{0.5cm}   C_{m,n} = \frac{\alpha}{\pi} \left[\frac{\sin q_F(m-n)}{\sin\alpha(m-n)}
    -\frac{\sin q_F(m+n)}{\sin\alpha(m+n)}\right]\,,\quad \quad \quad \alpha= \pi/2(L+1) \, ,
\label{corr_chain}
\end{equation}
where also $q_F= 2\alpha(K+1/2)$ if the $K$ momenta with $k=1,2,\dots,K$ are occupied.

Note that the Fermi momentum $q_F$ used here
(as in \cite{Fagotti/Calabrese11}) lies between the last occupied and the first unoccupied $q$-value,
while $\alpha$ is half the spacing of the momenta. As can be seen, the two expressions for $C_{m,n}$
are closely related, with the chain formula given by the ring expression minus an ``image term'' where
$n \rightarrow -n$. Only the values of $\alpha$ in the two cases are different. However, if one
considers a ring with $2L+2=2(L+1)$ sites and a chain with $L$ sites, they become exactly equal.
As dicussed below, this allows to ``embed'' the chain problem into a ring \cite{Fagotti/Calabrese11}.

We divide the system in two parts and consider as subsystem the one with sites $i=1,2,\dots,\ell$.
The corresponding entanglement Hamiltonian is then \cite{Peschel03,Peschel/Eisler09}
\begin{equation}
\mathcal{H}=  \sum_{i,j=1}^{\ell} \, H_{i,j} c^{\dag}_i c_j \, ,
\label{ent_ham}
\end{equation}
and the matrix $H_{i,j}$ is given in terms of the eigenfunctions $\phi_k(i)$ and 
eigenvalues $\zeta_k$ of the correlation matrix $C_{i,j}$ restricted to the subsystem
\begin{equation}
H_{i,j}= \sum_{k=1}^{\ell}\,\phi_k(i)\; \varepsilon_k\; \phi_k(j) \, ,
\label{spectral_H}
\end{equation}
where $\varepsilon_k = \ln [(1-\zeta_k)/\zeta_k]$.

The relation between chain and ring can be used as follows. Consider
a subsystem in a ring with $2\ell+1$ sites numbered $i=-\ell,\dots,0, \dots,\ell$. The eigenfunctions
of $C_{i,j}$ are either symmetric or antisymmetric under reflection of $i$, and it is easy to see
that the antisymmetric ones are the eigenfunctions (with the same eigenvalue) of the chain correlation
matrix \eqref{corr_chain} in the subsystem with the $\ell$ sites $i=1,2,\dots,\ell$. One only has to multiply
them by $\sqrt 2$ to obtain the normalization in the chain.
Thus, in principle, one has only to solve the ring problem. In the limit $L \rightarrow \infty$,
one obtains the connection between subsystems in infinite and at the end of semi-infinite chains
noted in \cite{Eisler/Peschel13}.

 \section{Commuting operator}

 A remarkable property of the hopping model is that the correlation matrix $C$, and therefore also
the matrix $H$, commute with a tridiagonal matrix $T$ of the form
\begin{equation}
   T =  
    \left(  \begin{array} {ccccc}
      d_1 & t_1   &  &   & \\  
       t_1 & d_2 & t_2  &  &\\
       & t_2 & d_3 & t_3  &  \\
        & & \ddots & \ddots & \\
        & & & t_{\ell-1} & d_{\ell}
  \end{array}  \right) .
  \label{tridiagonal1}
\end{equation}
The significance of this feature is twofold. Mathematically, the Jacobi-type matrix $T$ has a
simple (i.e. non-degenerate) spectrum and the eigenvectors can be obtained without numerical problems,
in contrast to the situation with $C$. Physically, it has the structure of $\hat H$ in \eqref{hamhop}
and describes a hopping model in the subsystem with spatially varying parameters, if one forms 
\begin{equation}
\mathcal{T}=  \sum_{i,j=1}^{\ell} \, T_{i,j} c^{\dag}_i c_j \, .
\label{T_ham}
\end{equation}
This operator then commutes with the entanglement Hamiltonian, $[\mathcal{H},\mathcal{T}]=0$.

The property is known to hold in the following cases: an interval in an infinite chain
\cite{Slepian78,Peschel04}, an interval at the end of a semi-infinite chain \cite{Peschel04}
and an interval in a finite ring \cite{Grünbaum81,Xu/Chamzas84}. In the following, we prove
it also for an interval at the end of a finite chain.

 Since $C$ and $T$ are symmetric, their commutator can be written as
\begin{equation}
[C,T]_{i,j}= (CT)_{i,j}-(CT)_{j,i} \, .
\label{comm1}
\end{equation}
Inserting the explicit form 
\begin{equation}
T_{k,j}= t_{j-1}\delta_{k,j-1}+d_j\delta_{k,j}+t_{j}\delta_{k,j+1} \, ,
\label{formulaT}
\end{equation}
writing $C=C^{-}-C^{+}$ according to Eq. \eqref{corr_chain}
and using the symmetry of the matrices $C^{\pm}$ as
well as the translational invariance of $C^{-}$, one finds
\begin{eqnarray}
  [C^{+},T]_{i,j} &=& C^{+}_{i,j-1}(t_{j-1}-t_{i-1})+ C^{+}_{i,j+1}(t_{j}-t_{i})+ C^{+}_{i,j}(d_{j}-d_{i}) \, ,
\label{comm_plus}
\end{eqnarray}
and
\begin{equation}
[C^{-},T]_{i,j}= C^{-}_{i,j-1}(t_{j-1}-t_{i})+ C^{-}_{i,j+1}(t_{j}-t_{i-1})+ C^{-}_{i,j}(d_{j}-d_{i}) \, .
\label{comm_minus}
\end{equation}
One sees that only differences of the $t$'s and $d$'s appear, which leaves two additive constants free.
If the factors multiplying the $C$'s can be made to cancel the denominators, a much simpler expression
will result. This can actually be achieved by choosing them as
\begin{equation}
t_i = \cos 2\alpha (i+\frac{1}{2})\,, \qquad d_i = d\,\cos 2\alpha i \, ,
\label{choice}
\end{equation}
and using the difference formula for the cosine. Then, for example, the first term in \eqref{comm_minus}
becomes, up to the factor $\alpha/\pi$, 
\begin{eqnarray}
  C^{-}_{i,j-1}(t_{j-1}-t_{i}) &=& \frac{\sin q_F(i-j+1)}{\sin\alpha(i-j+1)}\,
  2\sin\alpha(i+j)\sin\alpha(i-j+1) \nonumber\\
  &=& 2 \sin q_F(i-j+1) \sin\alpha(i+j) \, .
\label{example}
\end{eqnarray}
Somewhat amazingly, this works for \emph{all} terms in \eqref{comm_plus} and \eqref{comm_minus} and one
ends up with
\begin{equation}
[C,T]_{i,j}= \frac{\alpha}{\pi} \,(4\cos q_F +2d)\,[\sin q_F(i-j)\sin\alpha(i+j)-\sin q_F(i+j)\sin\alpha(i-j)] \, .
\label{comm_final}
\end{equation}
This expression then vanishes if one chooses $d= -2\cos q_F$, and this holds for all sites $i,j$
in the interior. On the left boundary ($i=1$ or $j=1$), the coefficient $t_{0}$ appears which is
not contained in $T$, but it is multiplied by $C_{i,0}$ or $C_{0,j}$ which vanish. On the right boundary
($i=\ell$ or $j=\ell$), the equations \eqref{comm_plus} and \eqref{comm_minus} contain the coefficient
$t_{\ell}$, which also does not appear in $T$. But it can be given the value zero by subtracting a constant
from all $t_i$. Subtracting also a convenient constant from all $d_i$, the final expressions are
%
\begin{equation}
\begin{split}
 t_i &=  \cos \alpha(2i+1)-\cos \alpha(2\ell+1) \, , \\
 d_i &= -2 \cos q_F\left(\cos 2\alpha i -\cos 2\alpha \ell \right) .
\label{coeff_final}
\end{split}
\end{equation}
Qualitatively, both quantities have their maxima at the left boundary, decrease towards the interior
and vanish at
$i=\ell$, i.e. at the right end of the subsystem. The diagonal terms are zero for half filling,
$q_F=\pi/2$, which is possible for even $L$. Then, for a half-chain, $\ell=L/2$, the hopping
elements take a simple form, if the numbering $r=\ell-i$ starts from the middle of the chain
\begin{equation}
\tilde t_r = \sin(2\alpha r)\, , \quad \quad r=1,2,\dots,  \ell-1 \, .
\label{half_half}
\end{equation}
By keeping $\ell$ fixed and letting $L \rightarrow \infty$,
one can recover the result for an interval at the end of a half-infinite chain. Then
$\alpha \rightarrow 0$ and the cosine functions can be expanded near their maxima to obtain the
parabolic laws
%
\begin{equation}
\begin{split}
 t_i &=  2\alpha^2 (\ell+1+i)(\ell-i) \, , \\
 d_i &= -4\alpha^2 \cos q_F (\ell+i)(\ell-i) \, .
\label{coeff_asymp}
\end{split}
\end{equation}

The general formula \eqref{coeff_final} also gives the result of Gr\"unbaum \cite{Grünbaum81}
for the ring in the particular case of a subsystem between $i=-\ell$ and $i=\ell$. 
On a ring, where $C=C^{-}$, one can shift the arguments of the cosines, i.e. the position of the
subsystem. If it lies between $i=1$ and $i=\ell$, one obtains by shifting, subtracting different
constants in \eqref{coeff_final} and converting to sines
%
\begin{equation}
\begin{split}
 t_i &=  \sin \alpha i \,\sin \alpha(\ell-i) \, , \\
 d_i &= -2\cos{q_F} \sin \alpha(i-1) \,\sin \alpha (\ell-i) \, .
  \label{coeff_ring} 
\end{split}
\end{equation}
 %
 
 Up to a prefactor, which will be included later, all these expressions correspond exactly to
those obtained for the entanglement Hamiltonian within conformal field theory \cite{Cardy/Tonni16}.
In that case, the trigonometric functions arise from the specific conformal mappings, whereas
for $T$, as seen above, their origin lies in the functional form of the correlation matrix.

\section{Entanglement Hamiltonian}

We now turn to the properties of $\mathcal{H}$ as obtained from numerical calculations.
In order to avoid eigenvalues $\zeta=0$ or $\zeta=1$ in $C$, which lead to $\varepsilon=\pm\infty$,
the total particle number $N$ should satisfy $\ell \le N \le L-\ell$ \cite{Xu/Chamzas84}.
If $L$ is even and the subsystem is half the total one, $\ell=L/2$, this forces also $N=L/2$,
i.e. half filling, and in this section we will always consider this case. To find the large
eigenvalues $\varepsilon_k$ reliably, we worked with about $3\ell/2$ digits in the diagonalization of
$C$ which is possible using \emph{Mathematica} as in \cite{Eisler/Peschel17}.

\begin{figure}[htb]
  \center
  \includegraphics[width=0.49\textwidth]{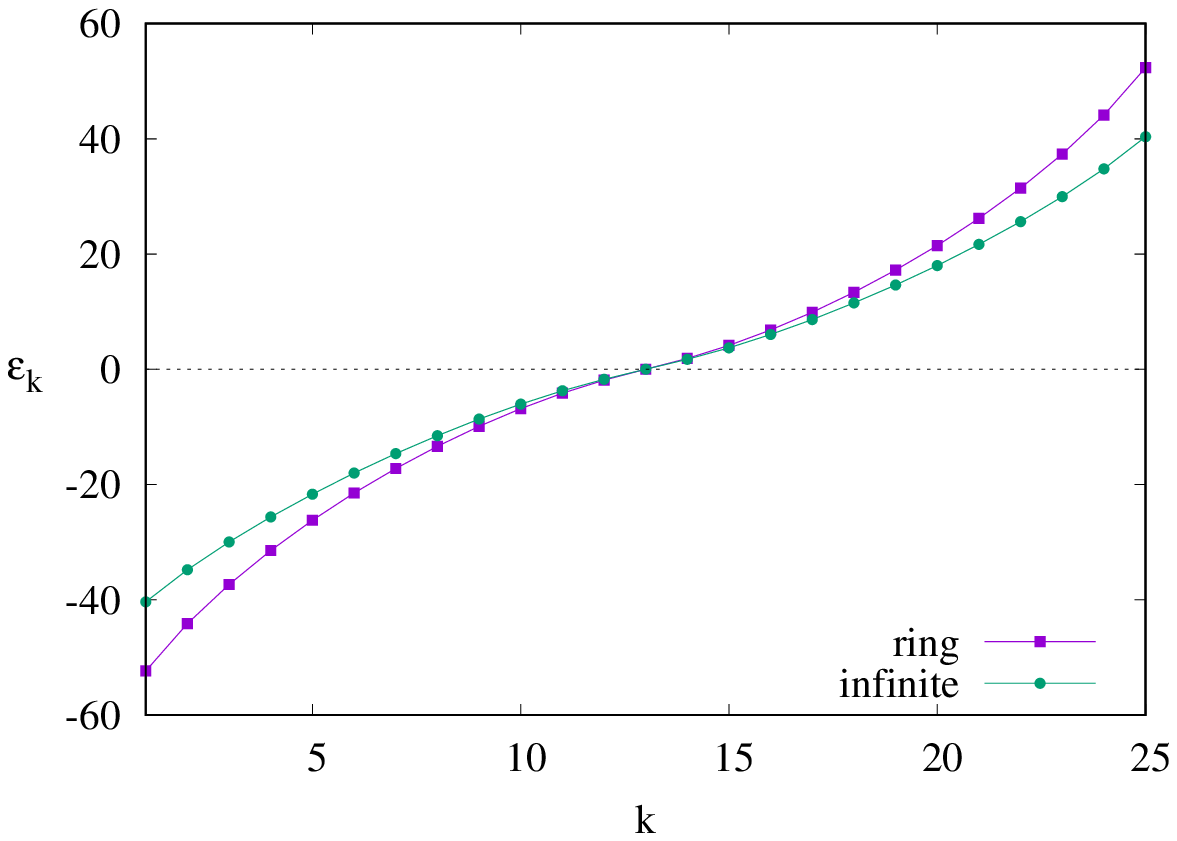}
  \includegraphics[width=0.49\textwidth]{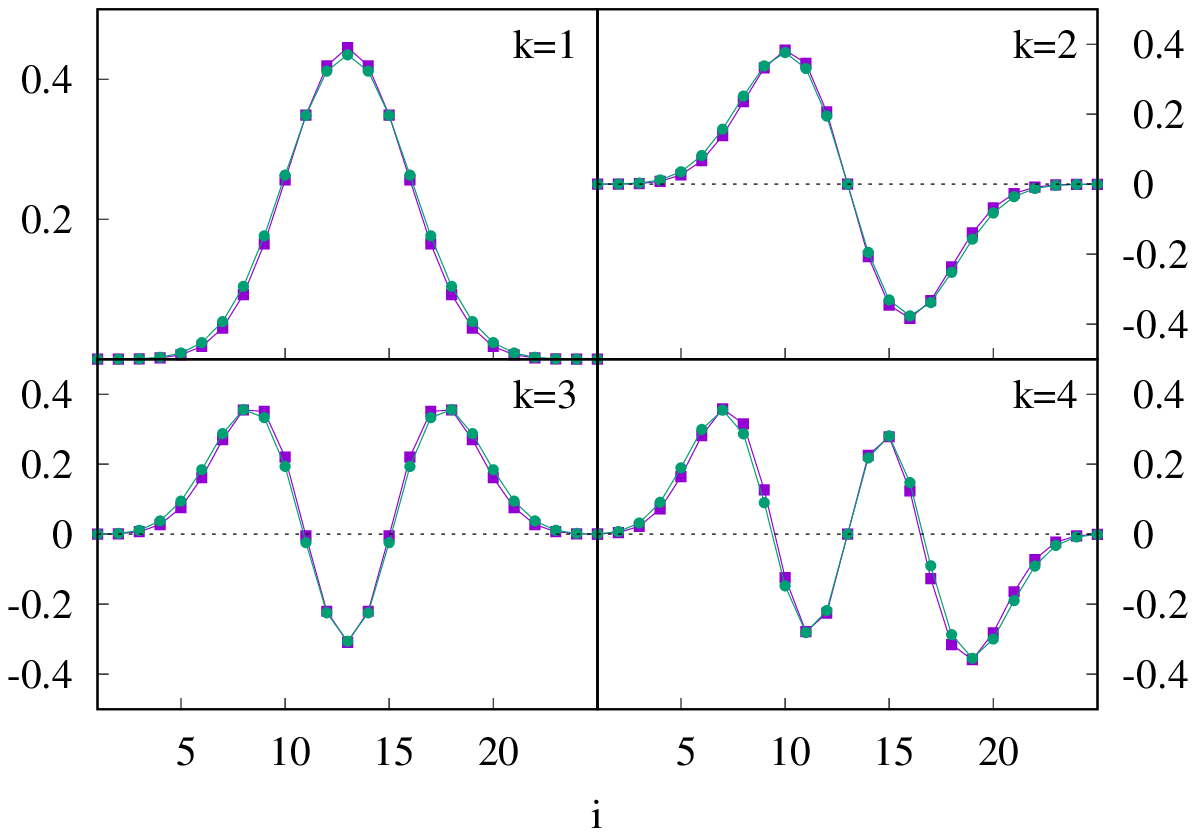} 
\caption{Left: Single-particle eigenvalues $\varepsilon_k$ for a subsystem of $\ell=25$ sites in
  a half-filled ring of $L=50$ sites and in an infinite system. Right: Eigenvectors $\phi_k(i)$
  for the four smallest $k$ in both cases.}
\label{fig:eps}
\end{figure}

In Fig. \ref{fig:eps}, the $\varepsilon_k$ for an interval of $\ell=25$ sites in a ring with $L=50$
are shown, together with those in an infinite system. Due to half filling, the spectra are symmetric
about zero, and the odd number of sites gives one vanishing eigenvalue. One sees that both spectra are 
rather similar, but the absolute values of the $\varepsilon_k$ are larger for the ring, in particular
at the upper and lower end. This leads to a smaller entanglement entropy, and the difference is
well described by the conformal result
\eq{
S_{inf}-S_{ring} = \frac{1}{3} \ln \frac {\pi}{2} \approx 0.1505 \, ,
}
which follows from the general expression
$S_{CFT}=1/3 \ln \left(L/\pi \sin \pi \ell/L \right) + k$ \cite{Calabrese/Cardy04}.
Note that this smaller entanglement is in contrast to the decay of the correlations across the
subsystem which is slower for the ring, because $C_{1,j}$ has to increase again beyond $j=L/2$.

%
%

In spite of the clear difference of the eigenvalues, the eigenvectors $\phi_k(i)$ in both cases
look essentially identical. This is illustrated for the four lowest (most negative) $\varepsilon_k$ 
on the right of Fig. \ref{fig:eps}. The feature remains valid all the way through the spectrum and
actually even improves as one moves into the centre. We have also checked it for larger $L$.
As a result, one can obtain the small $\varepsilon_k$ of the finite ring rather well
by taking the eigenvectors $\phi_k^{\infty}$ of the infinite system and forming
$\zeta'_k=\langle \phi_k^{\infty}| C^{\mathrm{ring}}| \phi_k^{\infty} \rangle$. For the large $\varepsilon_k$,
which come from $\zeta'_k$ very close to zero or one, the tiny eigenvector differences matter and
the resulting values lie close to those of the infinite system.

%
\begin{figure}[htb]
\center
\includegraphics[width=0.49\textwidth]{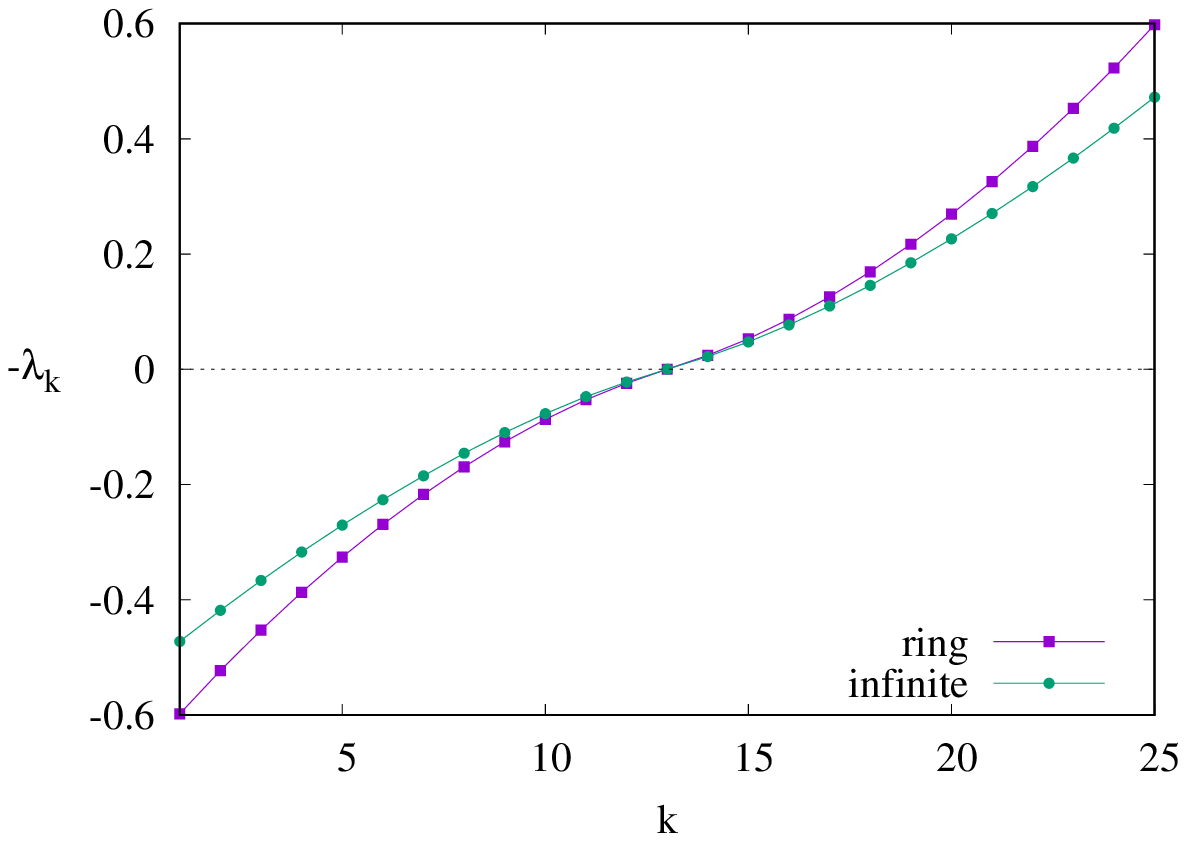}
\includegraphics[width=0.49\textwidth]{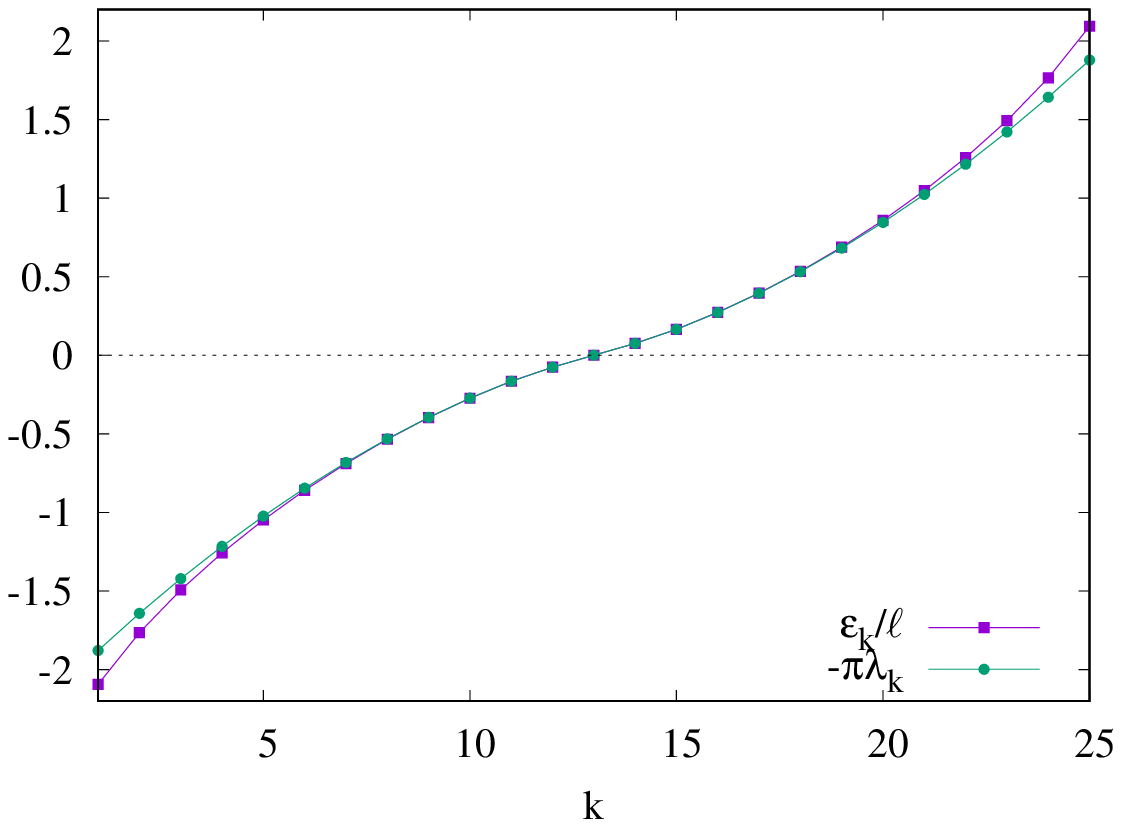}
\caption{Left: Eigenvalues $\lambda_k$ of $T$ for ring and infinite system. The parameters are the
         same as in Fig. \ref{fig:eps}.
         Right: Comparison of $\varepsilon_k/\ell$ and $\pi\lambda_k$ for the ring.}
\label{fig:lam}
\end{figure}
%

Next, we compare the eigenvalues $\lambda_k$  of the commuting matrices $T$
for the ring and the infinite system and the same parameters. Due to the half filling, the
diagonal terms in $T$ are zero. To be consistent with the CFT formulae \cite{Cardy/Tonni16}
in the continuum limit and to normalize with respect to $\ell$, we divide the $t_i$ in
\eqref{coeff_ring} (and thus $T$) by a factor and write 
\begin{equation}
  t_i =  \frac{\sin \alpha i \,\sin \alpha(\ell-i)}{\alpha\ell\sin\alpha\ell}\,,
          \quad \quad \quad \alpha= \pi/L \, .
\label{coeff_ring_rescaled}
\end{equation}
Then $t_i \simeq i/\ell$ for small $i$ independent of $L$. For $\ell=L/2$, the prefactor is simply $2/\pi$
and $t_i$ becomes
\begin{equation}
 t_i = \frac{1}{\pi} \sin \frac{\pi i}{\ell}\,,
\label{coeff_halfring}
\end{equation}
while for $L \rightarrow \infty$, one obtains the standard expression of the infinite system 
\begin{equation}
 t_i \simeq \bar t_i = \frac{i}{\ell} \left( 1-\frac{i}{\ell} \right).
\label{coeff_inf}
\end{equation}

The resulting $\lambda_k$ are shown in
Fig. \ref{fig:lam} on the left and look very similar to the $\varepsilon_k$. In particular,
the absolute values for the ring are again larger. In the infinite system, $\varepsilon_k/\ell$
and $-\pi\lambda_k$ are known to lie rather close to each other \cite{Eisler/Peschel17}, and the right
part of Fig. \ref{fig:lam} demonstrates that this also holds for the ring. The connection between the
two quantities will be further analyzed in Section 5.

\begin{figure}[htb]
  \center
  \includegraphics[width=0.49\textwidth]{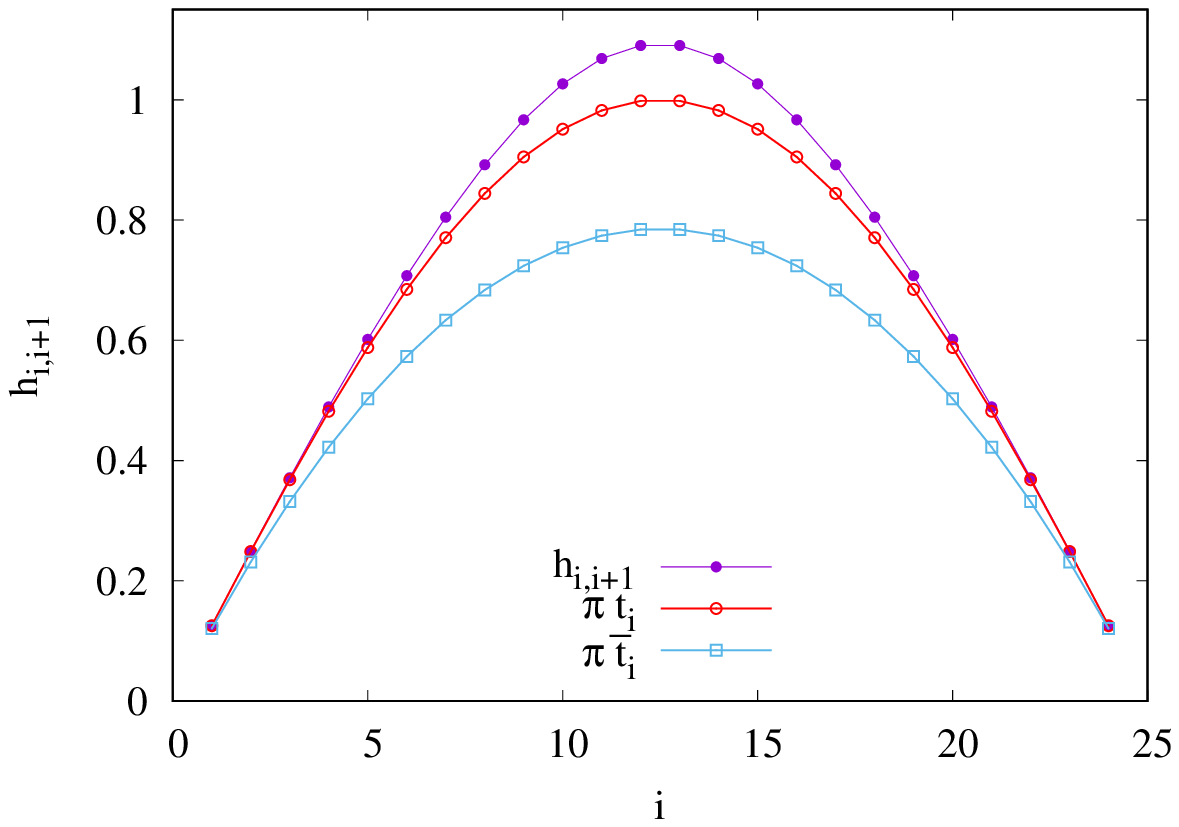}
\includegraphics[width=0.49\textwidth]{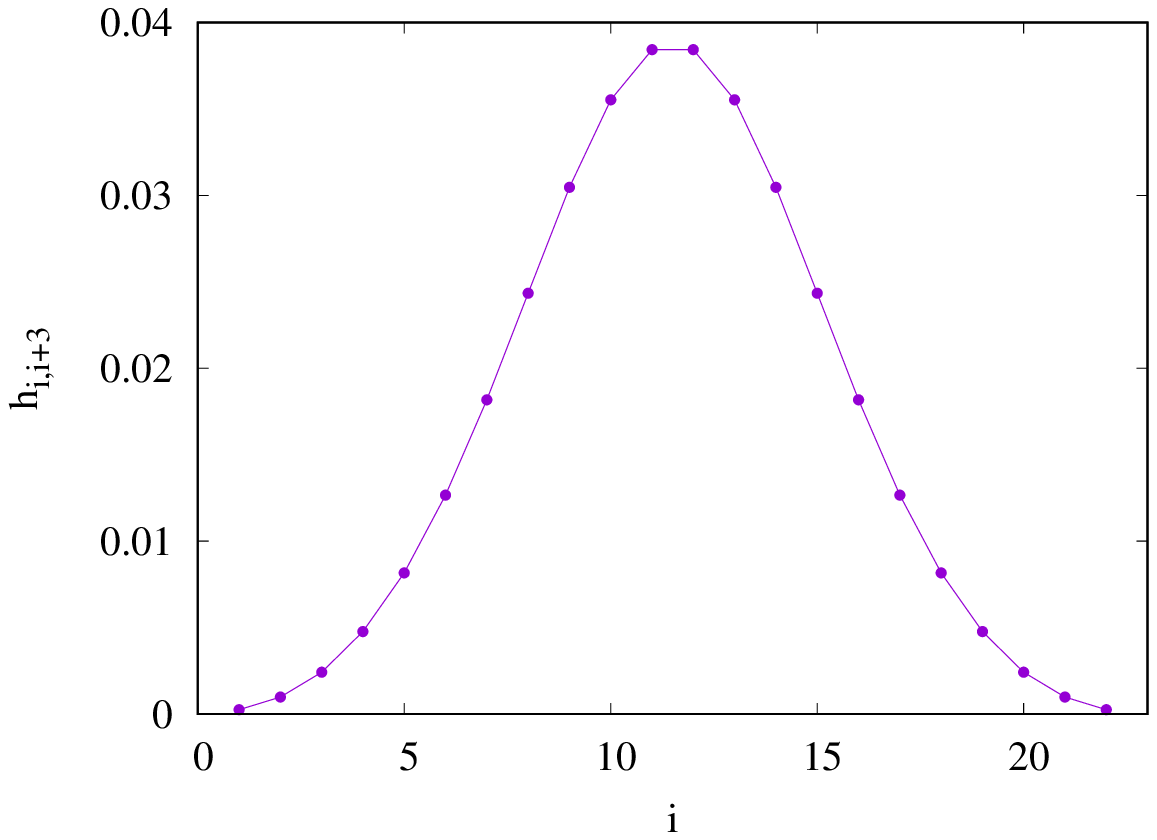}
\caption{Entanglement Hamiltonian for an interval of $\ell=25$ sites in a ring with $L=50$.
  Left: Nearest-neighbour hopping (top curve), together with $\pi t_i$ for the ring (middle) and
  $\pi \bar t_i$ for the infinite system (bottom). Right: Hopping to third neighbours. Note the
  different vertical scales.}
\label{fig:enthp}
\end{figure}

Results for the normalized matrix elements
\begin{equation}
h_{i,j}= -H_{i,j}/\ell\,,
\label{H_normal}
\end{equation}
whith $H_{i,j}$ given by  \eqref{spectral_H}, are shown in Fig. \ref{fig:enthp}. On the left, the
nearest-neighbour hopping $h_{i,i+1}$ is plotted (maximal value 1.09) together
with the corresponding quantity $\pi t_i$ according to \eqref{coeff_halfring} which is seen
to lie somewhat below it (maximal value 1) and the parabolic law $\pi \bar t_i$ of the infinite
system \eqref{coeff_inf} which again lies lower (maximal value $\pi/4$).
On the right, the hopping $h_{i,i+3}$ is plotted, which has no counterpart in $T$, 
is smaller by a factor of about $25$ and increases with a higher power near the boundaries.

%
%

\begin{figure}[htb]
\center
\includegraphics[width=0.49\textwidth]{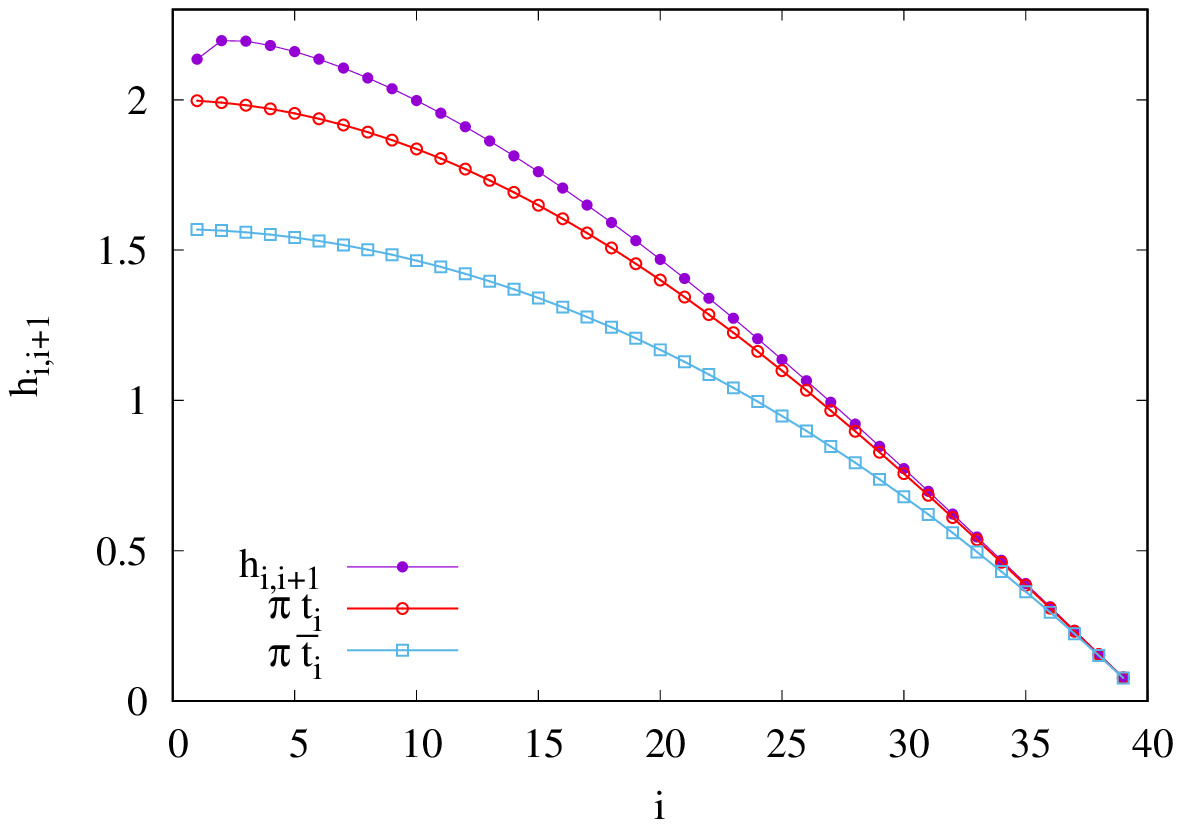}
\includegraphics[width=0.49\textwidth]{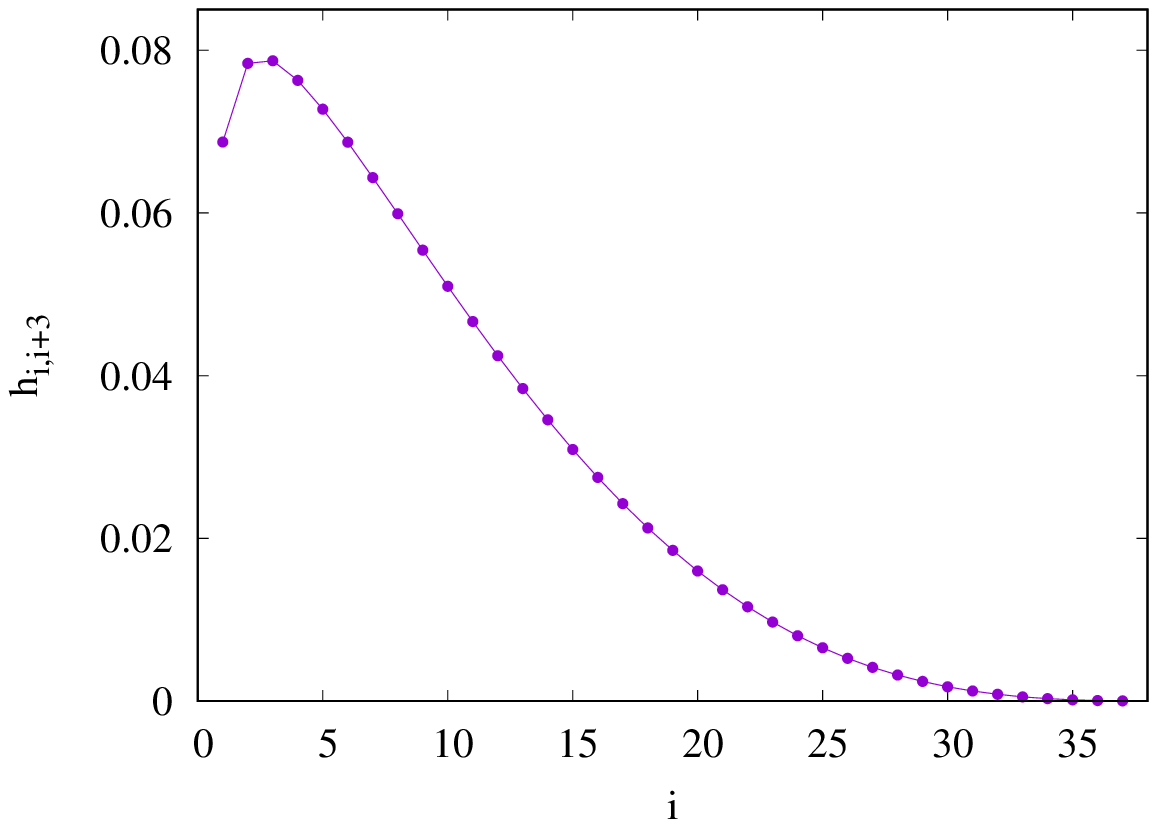}
\caption{Entanglement Hamiltonian for $\ell=40$ sites at the left end of a chain with $L=80$ sites.
        Left: Nearest-neighbour hopping (top curve) together
        with $\pi t_i$ for the chain (middle) and $\pi \bar t_i$ for the half-infinite
        system (bottom). Right: Third-neighbour hopping. Note the different
        vertical scales.}
\label{fig:entho}
\end{figure}

Finally, we turn to the open chain. According to the remarks in Section 2, the previous calculations
also give the result for a subsystem of $\ell=12$ sites at the end of a chain with $L=24$. However,
this is rather small, and we present in Fig. \ref{fig:entho} the results of a calculation for $\ell=40$.
One sees that the nearest-neighbour hopping increases from the centre towards the free boundary with
a maximum about twice the value for the ring. While the overall shape is roughly that of the right part in
the ring, there is one conspicuous difference: before one reaches the free boundary, there is a break
in the curve and the last value decreases. This feature also appears in the hopping to the
third neighbours shown on the right and is even stronger there. It comes from the eigenvectors
$\phi_k(i)$, which all vanish at $i=0$ and thus approach zero already inside the subsystem.
However, the effect depends on the magnitude of the largest eigen\emph{values} and does not appear if $\ell$
is too small. It also does not appear in $t_i$ which can be written as $H_{i,i+1}$ in \eqref{spectral_H}
with $\varepsilon_k$ replaced with $\lambda_k$. One should mention that a decrease of the hopping near
the boundary was also found in the rainbow chain, but there it takes place rather gradually
\cite{Tonni/Laguna/Sierra17}. 

%
%
 For the comparison on the left, $t_i$ in \eqref{coeff_final} was rescaled to read 
\eq{
t_i = \frac{\cos (2i+1)\alpha - \cos(2\ell+1)\alpha}
       {(2\ell+1)\alpha \sin (2\ell+1)\alpha} \,,\quad \quad \quad \alpha= \pi/2(L+1) \, .
\label{coeff_chain_rescaled}
}
For the half-chain, this gives again a factor of $2/\pi$ and
%
\begin{equation}
 t_i = \frac{2}{\pi} \sin \frac{\pi (\ell-i)}{2\ell+1} \, .
\label{coeff_halfchain}
\end{equation}
Note that in \cite{Cardy/Tonni16,Tonni/Laguna/Sierra17} the subsystem is at the right end of the chain,
which leads to an interchange of sines and cosines in the formula for the weight
factor.

As to the scaling behaviour, one finds that the $h_{i,j}$ collapse very well for different sizes, 
if plotted against $i/\ell$. The only exception is the vicinity of the boundary in the case of the chain.

\section{Asymptotics}

In this section, we study in more detail the behaviour of the eigenvalues of $H$ and $T$
for half filling and large subsystems, i.e. for large $\ell$. We do this separately for
the low-lying eigenvalues, the maximal ones and for the overall spectra.

\subsection{Low-lying eigenvalues}

In the infinite system, the asymptotic analysis of Slepian \cite{Slepian78}
(see \cite{Eisler/Peschel13} for a summary), gives the low $\varepsilon_k$ as
solutions of the equation
\eq{
\frac{\pi}{2}\left(k-\frac{1}{2}-\frac{ q_F \ell}{\pi}\right)=
\frac{\varepsilon_k}{2\pi} \ln (2\ell\sin q_F) -\varphi\left(\frac{\varepsilon_k}{2\pi}\right) ,
\label{keps}}
where $\varphi(y)=\mathrm{arg}\,\Gamma(1/2+iy)$. This can be solved explicitly for small $\varepsilon_k$
by using the linear approximation $\varphi(y) \simeq \psi(1/2)\,y$ with $\psi(y)$ denoting the digamma
function and $\psi(1/2) \simeq -1.963$. As a result, one obtains a linear spacing of the lowest
eigenvalues with a logarithmic density of states
\eq{
\varepsilon_k \simeq \frac{\pi^2}{\ln (2\ell\sin q_F)-\psi(1/2)}\left(k-\frac{1}{2}-\frac{ q_F \ell}{\pi}\right) .
\label{keps2}}

Now, it was observed in \cite{Tonni/Laguna/Sierra17} that the \emph{first gap} (i.e. the smallest
positive $\varepsilon_k$) in the entanglement
spectrum for the rainbow chain is given by a formula as \eqref{keps2}, but with $\ell$ in the logarithm
replaced with the corresponding conformal length $\mathcal{L}$ (i.e. the effective width of the strip geometry
the CFT is mapped onto by proper conformal transformations).
This suggests to try the form \eqref{keps} as an ansatz for the \emph{complete} low-lying spectrum,
replacing $\ell$ with the chord length
\eq{
\mathcal{L}= \ell f(r)\,, \qquad f(r)=\frac{\sin \pi r}{\pi r}\,, \qquad r = \frac{\ell}{L}
\label{conf_l_ring}
}
for the ring and $2\mathcal{L}$ for the chain. Note that the logarithm $\ln(2\mathcal{L}\sin q_F)$
differs from the quantity $\ln (2\pi \bar n \mathcal{L})$ which appears in the equation
for the continuous ring \cite{Eisler/Peschel13}, where $\bar n=N/L$ denotes the density. However, in
the limit of small density, using $\bar n= q_F/\pi$ on the lattice, both expressions coincide.

\begin{figure}[htb]
\center
\includegraphics[width=0.49\textwidth]{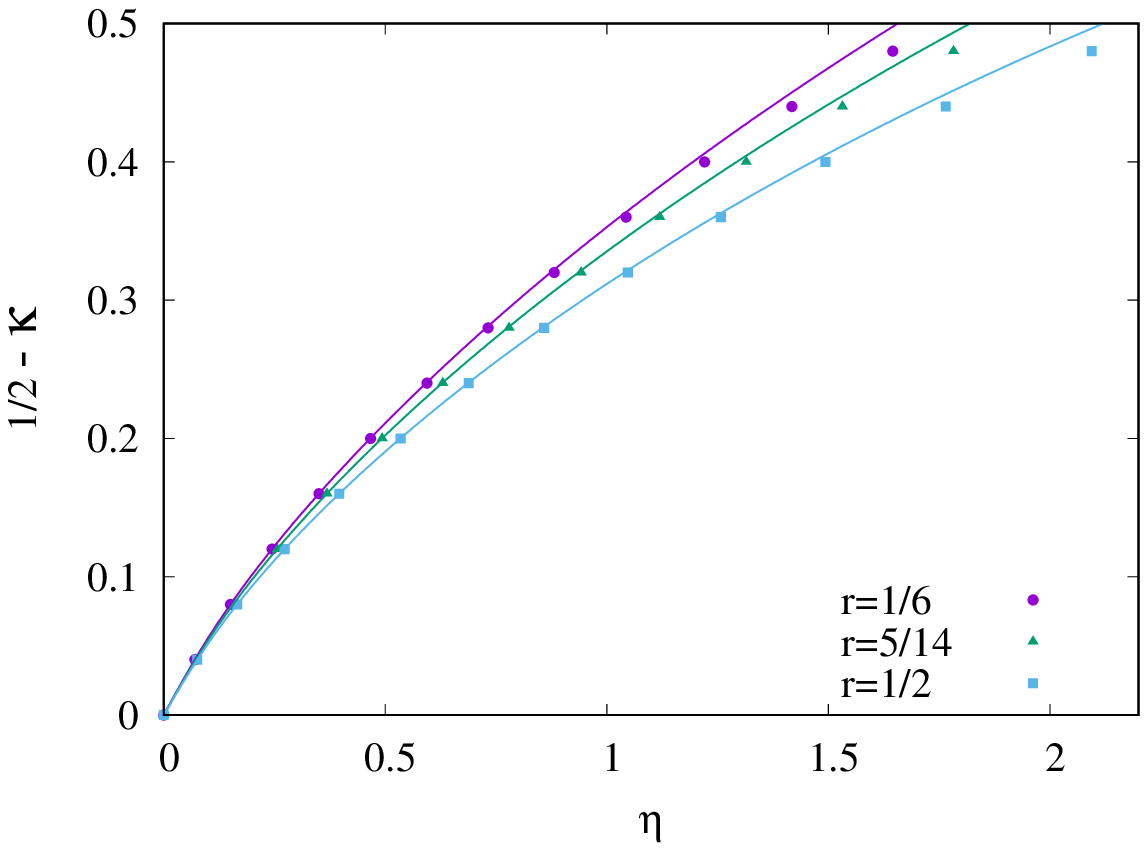}
\includegraphics[width=0.49\textwidth]{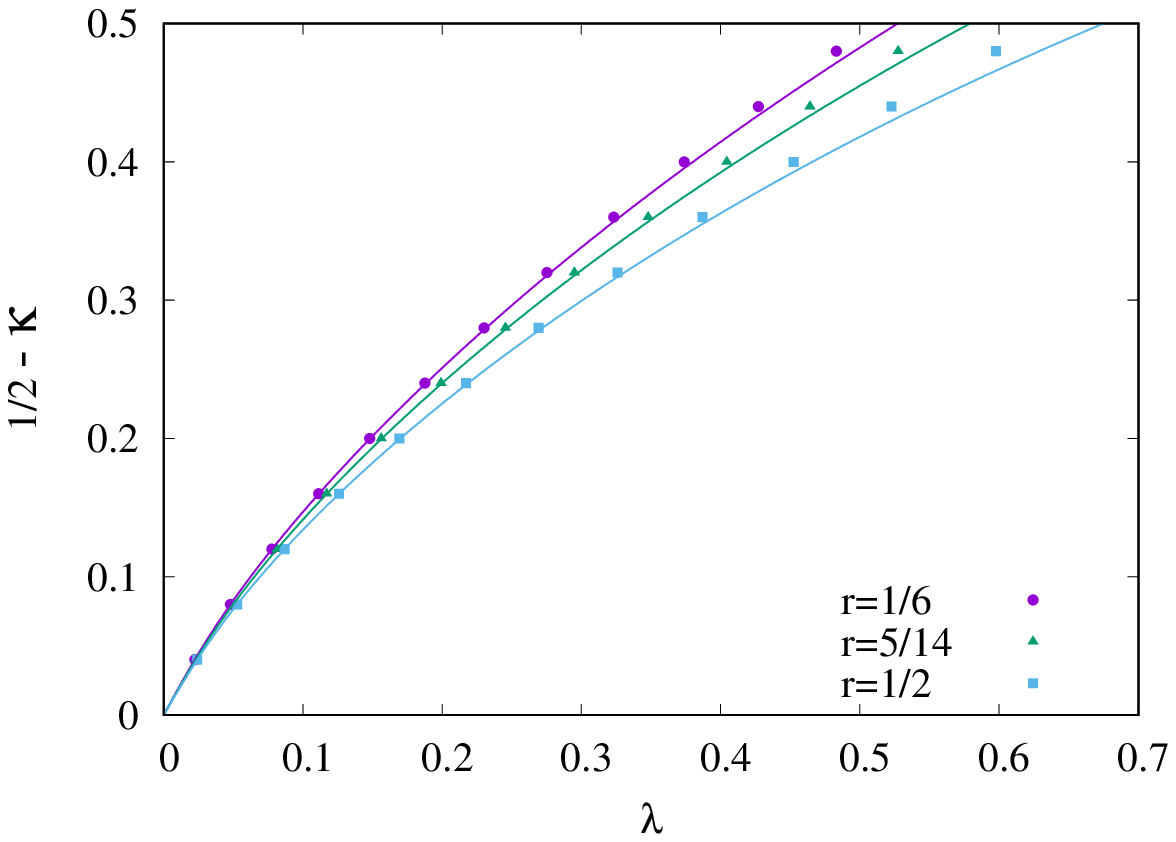}
\caption{Low-lying asymptotics for $\ell =25$ sites in rings with different ratios $r=\ell/L$.
  Shown is the scaled index $1/2-\kappa$  vs. $\eta$ (left) and vs. $\lambda$ (right).
  The symbols are the numerical data.}
\label{fig:lla}
\end{figure}

In terms of the scaled quantities 
\eq{
\eta
 = - \frac{\varepsilon_k}{\ell}, \qquad
\kappa = \frac{k-1/2}{\ell} ,
\label{def_eta_kappa}}
 the equation then becomes, for half filling, 
\eq{
\frac{1}{2}-\kappa = \frac{2}{\pi}
\left[\frac{\eta}{2\pi} \ln 2 \mathcal{L}-
\frac{1}{\ell}\,\varphi \left(\frac{\eta \ell}{2\pi}\right)\right].
\label{kappa_eta}}
The spectral parameter $\kappa$ varies in the range $0<\kappa<1$ with $\kappa=1/2$ corresponding
to the centre of the spectrum.
The solutions are shown in Fig. \ref{fig:lla} on the left for three different ratios $r$ together with
the real data for $\ell=25$. One sees that the formula describes them very well up to the middle of
the spectrum
($\eta \simeq 1$) and correctly gives the increase of $\eta$ with $r$ noted already in Section 4. 
On the right, the results for $\lambda_k \equiv \lambda$ are plotted, with the curves given by the
linear approximation $\lambda = \eta /\pi$. Again, this gives a rather good desription of the data.
Thus the ring spectra follow from Slepian's formula by a simple rescaling within the logarithm.

%
%
%
%
We note that, for large $\eta\ell/2\pi$, the expression \eqref{kappa_eta} simplifies
because one can write
\eq{
  \varphi(y) = \mathrm{Im\,} \ln \Gamma(1/2+iy) \simeq y\, (\ln y -1) \, ,
\label{phi_large}  
}
which leads to
\eq{
\frac{1}{2}-\kappa \simeq \frac{\eta}{\pi^2} \left(1-\ln \frac{\eta}{4\pi f(r)}\right) ,
\label{kappa_large} }
and shows the correction to the linear term and its functional dependence on $r$ explicitly.
In particular, the spectrum $\eta$ is nonanalytic around the centre,
signalling the transition to a logarithmic density of states, which is responsible for the
$\ln \mathcal{L}$ scaling of the entanglement.

\subsection{Maximal eigenvalues}

\begin{figure}[htb]
\center
\includegraphics[width=0.49\textwidth]{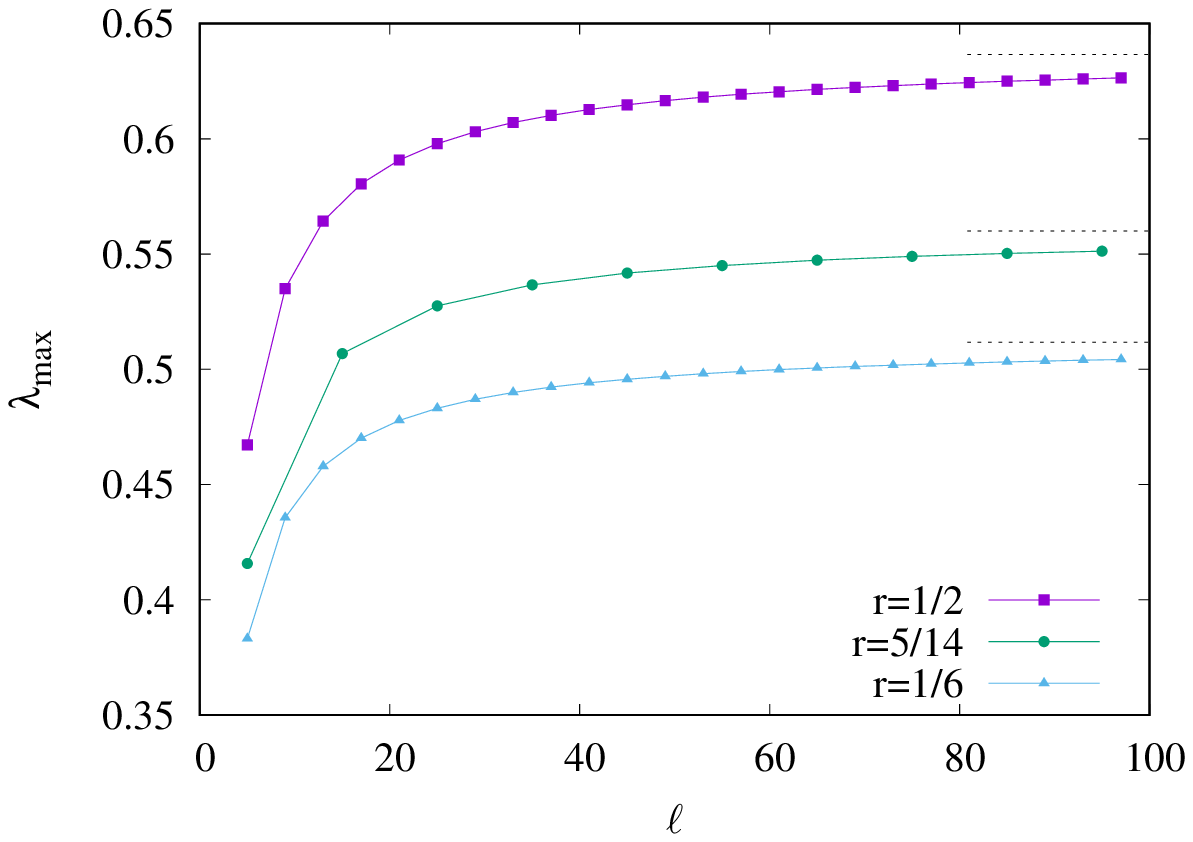}
\includegraphics[width=0.49\textwidth]{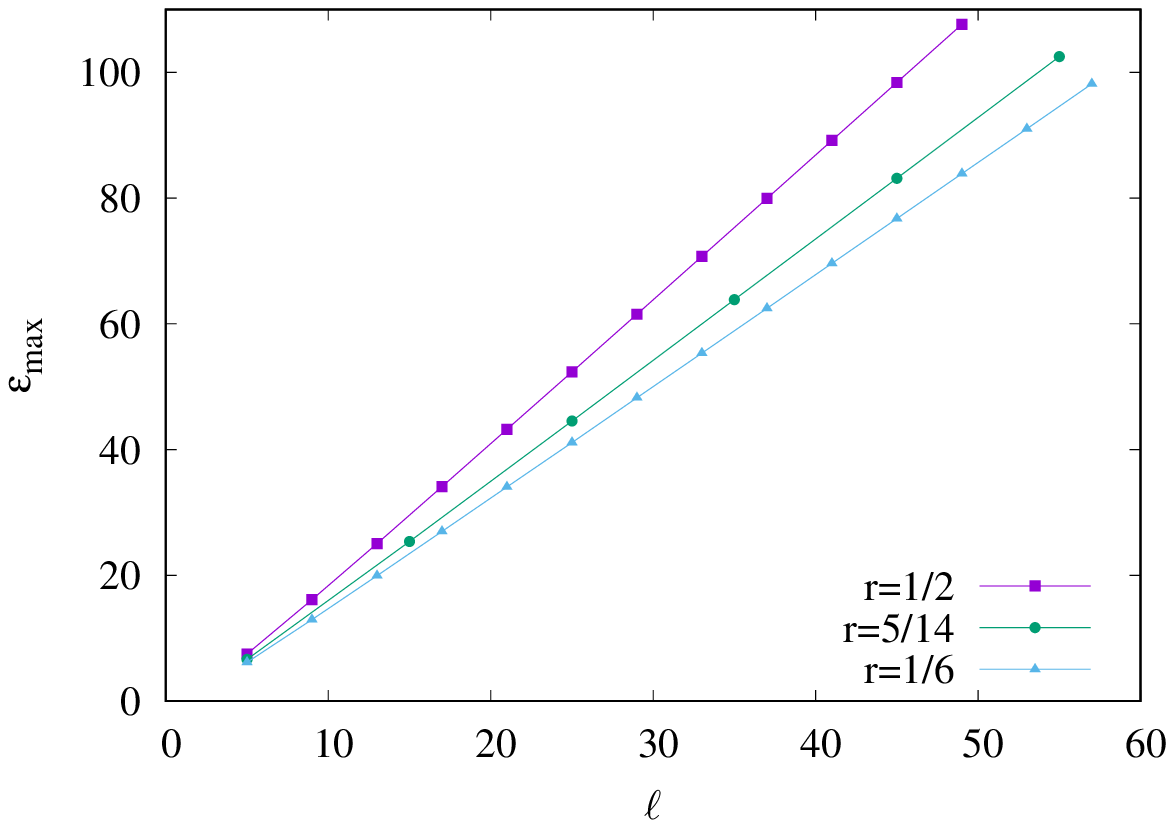}
\caption{Maximal eigenvalues for rings as functions of $\ell$ for
  different ratios $r=\ell/L$. Left: $\lambda_{\mathrm{max}}$. The dash-dotted lines are the
  values \eqref{lambda_max3}.
  Right: $\varepsilon_{\mathrm{max}}$. The slopes are, from top to bottom,
  a = 2.333, 1.947, 1.797. } 
\label{fig:lambda_eps_max}
\end{figure}

We now turn to the scaling of the maximal eigenvalues. For $\lambda_{\mathrm{max}}$, the asymptotic limit is
rather easy to obtain, since the corresponding eigenfunction has Gaussian form,
see $k=1$ on the right of Fig. \ref{fig:eps}. Writing
\eq{
\lambda_{\mathrm{max}} = \sum_{i,j} \phi_1(i) T_{i,j} \phi_1(j)
\simeq 2\int_{0}^{1} \dd x \, t(x) \phi^2_1(x) \,,\qquad x=j/\ell\,,
\label{lambda_max1}
}
inserting $t(x)$ from \eqref{coeff_ring_rescaled} as well as $\phi_1(x)$
\eq{
t(x) = \frac{\sin \pi r x \sin \pi r(1-x)}{\pi r \sin \pi r}, \qquad
\phi_1(x)=\frac{\ee^{-\frac{(x-1/2)^2}{4\sigma^2}}}{(2\pi \sigma^2)^{1/4}} \, ,
}
and extending the integration to $\pm \infty$, since the width of the Gaussian is 
$\sigma \propto 1/\sqrt{\ell}$, one can evaluate \eqref{lambda_max1}
to obtain
\eq{
  \lambda_{\mathrm{max}} = \frac{\ee^{-2(\pi r \sigma)^2}-\cos \pi r}{\pi r \sin \pi r}.
  \label{lambda_max2}
}
Up to corrections of order $1/\ell$, the exponential function is equal to one. The
asymptotic value therefore is
\eq{
  \lambda_{\mathrm{max}} = \frac{\tan (\pi r/2)}{\pi r} ,
  \label{lambda_max3}
}
and lies between $2/\pi \simeq 0.637$ for $r=1/2$ and $1/2$ for $r=0$.
In Fig. \ref{fig:lambda_eps_max}, numerical data are shown and one sees that they approach this limit
(from below, in accordance with \eqref{lambda_max2}) as $\ell$ increases. 
%
%
%

To obtain $\zeta_{\mathrm{max}}$ and the resulting $\varepsilon_{\mathrm{max}}$ is much more difficult. 
Calculating $\zeta_{\mathrm{max}}$ as the expectation value $\langle \phi_1| \, C \, |\phi_1\rangle$
together with a continuum limit as in \eqref{lambda_max1} gives an
$\varepsilon_{\mathrm{max}}$ which is proportional to $\ell$, but with incorrect prefactor. The reason
has to be sought in the rapid oscillations of $C$. We therefore present only numerical
results on the right side of Fig. \ref{fig:lambda_eps_max}. They can be fitted with an ansatz
$\varepsilon_{\mathrm{max}} = a \ell + b \ln \ell + c$ as for the infinite case.
The asymptotic value $\eta_{\mathrm{max}}=a$ is given by the slope of these curves, 
which shows a slow increase with $r$ similarly to $\lambda_{\mathrm{max}}$.

\subsection{Relation between $\eta$ and $\lambda$}

The difference between the matrix $h$ in the entanglement Hamiltonian and the commuting $\pi T$ is
completely due to their different spectra, see Fig. \ref{fig:lam}. Therefore, the first step in finding a general relation
between both quantities, is to connect their spectra. In the infinite system, this was possible
with the help of an asymptotic integral expression for $\eta$ which led to the formula \cite{Eisler/Peschel17}
%
\eq{
\eta=\pi\lambda
\; _{3}F_{2}\left(\frac{1}{4},\frac{1}{2},\frac{3}{4};1,\frac{3}{2};[2\lambda]^2\right),
\label{etalam}}
with a generalized hypergeometric function $_{3}F_{2}$.

\begin{figure}[htb]
\center
\includegraphics[width=0.49\textwidth]{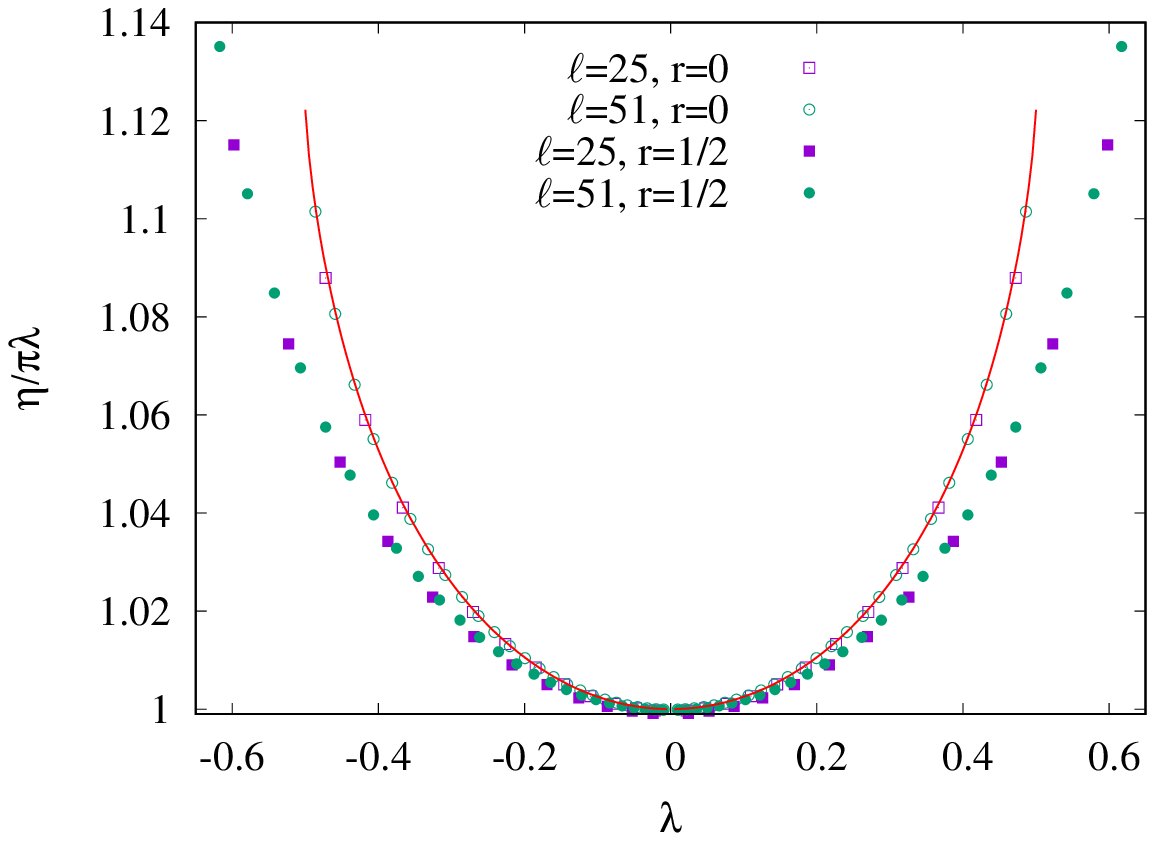}
\includegraphics[width=0.49\textwidth]{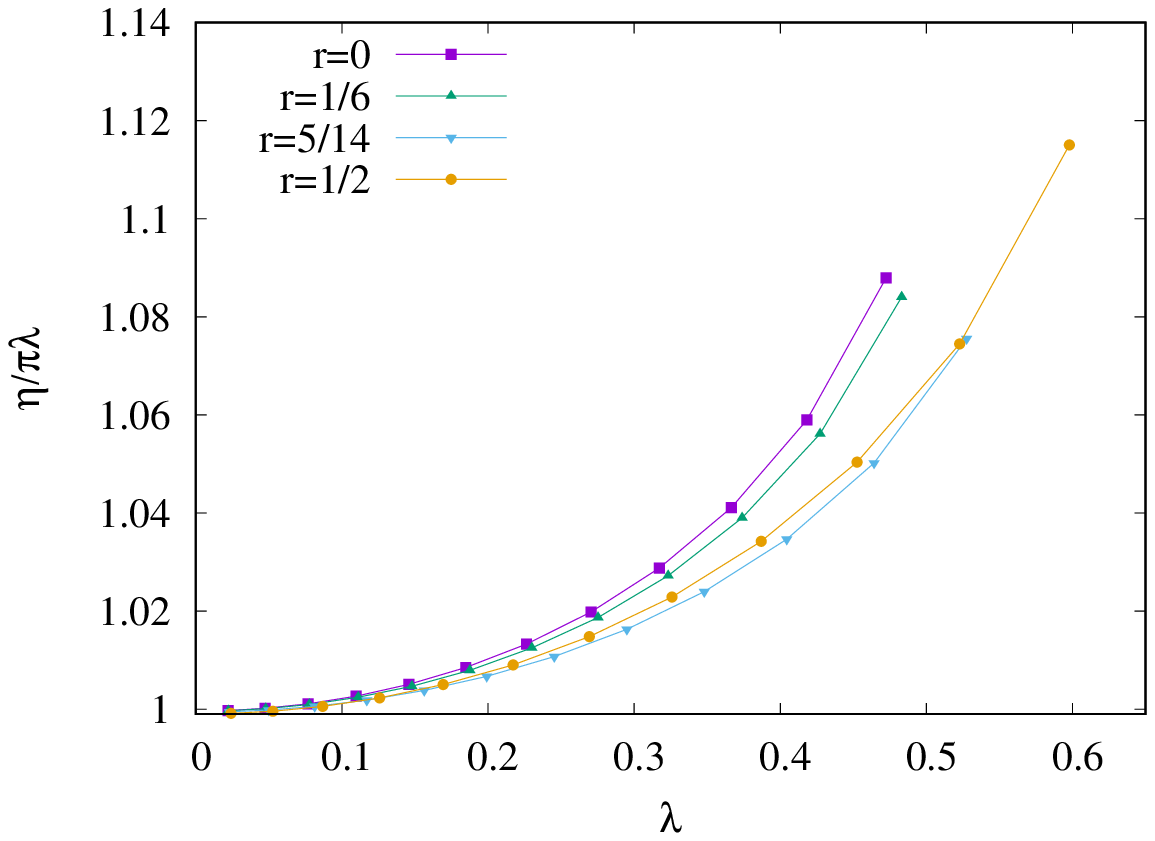}
\caption{Functional relation $\eta(\lambda)$. Shown is the ratio $\eta/\pi\lambda$.
  Left: half-ring and infinite chain for two values of $\ell$.
  The solid red line is the asymptotic $_3 F_2$ function in Eq. \eqref{etalam}.
  Right: The relation for $\ell=25$ and four different values of $r$.}
\label{fig:etalam}
\end{figure}

For finite rings or chains, the numerics show clearly, that there is an analogous relation between
these quantities which, however, depends on the ratio $r=\ell/L$. This is demonstrated on the left of
Fig. \ref{fig:etalam} for the half-ring with $r=1/2$ and compared to the infinite case $r=0$.
%
%
%

On the right, the $r$-dependence for the case $\ell=25$ is shown in more detail. One can see that,
starting from an infinite system, $\eta/\pi\lambda$ decreases at first, but finally, close to the
half-ring,
becomes larger again. This holds in particular for the maximal values. As a consequence, the ratio of
the hopping matrix elements $h_{i,i+1}/\pi t_i$ also shows such a non-monotonous behaviour as a
function of $r$. Another consequence is that the scaling function for $\eta/\pi\lambda$ cannot have
the simple form $F(\lambda/\lambda_{\mathrm{max}})$, but must depend on $r$ in a more complicated way.

In the infinite system, the asymptotic analysis of the eigenvalues $\varepsilon_k$ was actually
carried out by considering the reduced overlap kernel $A_{q,q'}$ between the occupied momentum
states via the equations \cite{Eisler/Peschel13}
\eq{
A_{q,q'} = \sum_{i=1}^{\ell} \Phi^*_q(i)\Phi_q(i), \qquad
\sum_{|q'| < q_F} A_{q,q'} \varphi_k(q') = \zeta_k \varphi_k(q), \qquad
\zeta_k = \sum_{|q|<q_F} |\varphi_k(q)|^2\,,
\label{overlap}
}
but using a commuting \emph{differential operator} to determine the eigenfunctions. This technique can
not be applied in finite rings, since the momenta are discrete and $A_{q,q'}$ is a matrix.
In fact, for the half-filled half-ring it has exactly the same form as $C_{i,j}$ with $i \rightarrow q$
and $q_F\rightarrow \ell/2$ and the commuting quantity is again the matrix $T$ but with momentum indices.
Therefore, one \emph{has} to treat a discrete problem. There are analogues of WKB techniques for
difference equations \cite{Dingle/Morgan67}, but we have not pursued this.

\section{Discussion}

We have studied free fermions hopping on finite rings and open chains and determined the
entanglement Hamiltonian from numerical calculations. This comprised the single-particle eigenvalue
spectrum as well as the functional form in real space with the focus on half-filled systems.
While these quantities depend on the aspect ratio $r=\ell/L$, the main feature was the
same as for subsystems in infinite systems: There is a difference between the properties of
$\mathcal{H}$ and the commuting operator $\mathcal{T}$ which is the lattice analogue of the
conformal entanglement Hamiltonian. We showed that such a $\mathcal{T}$ does exist also
for a subsystem at the end of an open chain and describe in the Appendix how to infer it
for the transverse Ising model.

In contrast to the infinite case, the number of analytical results was limited, because the
basic eigenvalue problem involves a finite matrix, and a simple continuum limit for the case of
large subsystems is in general not sufficient. Even for $T$, which is tridiagonal, the problem is
complicated, since the matrix elements are trigonometric functions. Nevertheless we could find
some analytical expressions for large $\ell$ for the small and large eigenvalues, but the general
scaling function relating $\eta$ and $\lambda$, and therefore $h$ and $T$, is still to be found.

In a broader context, the most intriguing feature is the existence of a simple commuting operator.
The treatment in Section 3 shows the mechanism in a very transparent way. One can also
prove \cite{Grünbaum81a} that a translationally invariant correlation
matrix commuting with a tridiagonal matrix with simple spectrum \emph{must} have the form
\eqref{corr_ring}. But $\mathcal{T}$ does not seem to fit into the scheme of conserved
quantities in the integrable spin models, and a physical explanation is still missing.
One should also mention that if one uses $\pi \ell\,\mathcal{T}$ in place of $\mathcal{H}$ in
the reduced density matrix, the resulting correlation function is not translationally invariant
on a ring, although the deviations are rather small. One can say that a nonlinear relation between
$\mathcal{H}$ and $\mathcal{T}$, as found in the infinite case, is necessary to restore this symmetry

\ack
We thank P. Calabrese and E. Tonni for helpful discussions.
V. E. acknowledges funding from the Austrian Science Fund (FWF) through Project No. P30616-N36.

\appendix

\section{Commuting operator for the TI chain}

The transverse-field Ising model is known to be closely related to the XX chain which in
turn is equivalent to the hopping model without site energies
\cite{Grady82,Peschel/Schotte84,Turban84,Igloi/Juhasz/Rieger00}. This allows to
find a commuting operator also for the TI chain without going through a separate calculation.

Consider a general open TI chain with Hamiltonian 
\begin{equation}
\mathcal{\hat H}_{TI} = -\frac{1}{2} \sum_{n=1}^L\,h_n \sigma_n^z
                   -\frac{1}{2} \sum_{n=1}^{L-1} \,\lambda_n\,\sigma_n^x \sigma_{n+1}^x \, .
\label{ham_ti}
\end{equation}
In terms of fermions, this becomes
\begin{eqnarray}
  \mathcal{\hat H}_{TI} &=& -\sum_{n=1}^L\,h_n \, c^{\dag}_n c_n 
                     -\frac{1}{2} \sum_{n=1}^{L-1} \,\lambda_n\,(c^{\dag}_n c_{n+1} +c^{\dag}_{n+1} c_{n}) 
                     -\frac{1}{2} \sum_{n=1}^{L-1} \,\lambda_n\,(c^{\dag}_n c^{\dag}_{n+1} +c_{n+1} c_{n})
                     \nonumber\\
                        &=& \sum_{n,m=1}^L\left[\hat A_{n,m}c^{\dag}_n c_m+\frac{1}{2}\hat B_{n,m}
                            (c^{\dag}_nc ^{\dag}_m+c_m c_n)\right] ,
\label{ham_tiferm}
\end{eqnarray}
and is diagonalized by a Bogoliubov transformation involving two functions $\Phi_q(n)$ and $\Psi_q(n)$.
They follow, together with the single-particle energies $\omega_q$, from the coupled
$(L\times L)$ systems \cite{LSM61}
\begin{equation}
(\hat A+\hat B)\,\Phi_q=\omega_q \Psi_q \, , \quad \quad
(\hat A-\hat B)\,\Psi_q=\omega_q \Phi_q \, .
\label{diag1}
\end{equation}
The eigenvalues come in pairs $(\omega_q,-\omega_q)$, and one can take e.g. the positive ones. Then the
ground state is the fermionic vacuum and the correlation matrix
$K_{m,n}=\langle (c^{\dag}_m+ c_m)(c^{\dag}_n-c_n) \rangle$ combining
$C_{m,n}=\langle c^{\dag}_mc_n \rangle$ and
$F_{m,n}=\langle c^{\dag}_mc^{\dag}_n \rangle$ is given by
\begin{equation}
  K_{m,n}= \sum_q\Phi_q(m)\Psi_q(n) \, .
\label{correl1}
\end{equation}
The entanglement Hamiltonian $\mathcal{H}$ for a subsystem, which has again the form \eqref{ham_tiferm}
but with matrices $A_{i,j}$ and $B_{i,j}$, can then be found from the restricted correlation matrix
$K_{i,j}$ and its transpose $K'_{i,j}$ by solving the coupled $(\ell \times \ell)$ systems
\begin{equation}
K'\,\phi_k=\tanh(\varepsilon_k /2) \,\psi_k \, , \quad \quad
K\,\psi_k=\tanh(\varepsilon_k /2)\, \phi_k \, ,
\label{diag2}
\end{equation}
and constructing $A$ and $B$ from $\phi_k,\psi_k$ and the single-particle eigenvalues $\varepsilon_k$
\cite{Peschel03}.

Now, it was pointed out in \cite{Igloi/Juhasz08} that by writing \eqref{diag1} as a single
$(2L\times 2L)$ system for the vector
\begin{equation}
\hat \chi_q=\left( \Phi_q(1),\Psi_q(1),\Phi_q(2),\Psi_q(2),\dots,\Psi_q(L) \right)/\sqrt2\,,
\label{def_chi}
\end{equation}
the matrix $\hat M$ in $\hat M\hat \chi_q=\omega_q \,\hat \chi_q$ has the form
\begin{equation}
   \hat M = - 
    \left(  \begin{array} {ccccc}
      0  & h_1  &  &   & \\  
       h_1 & 0 & \lambda_1  &  & \\
       & \lambda_1 & 0  & h_2  &  \\
         & & h_2 & 0 & \ddots \\
           & & & \ddots & \ddots \\
  \end{array}  \right),
  \label{tridiagonal3}
 \end{equation}
 and thus is the same as for an inhomogeneous hopping model with $2L$ sites
 and no site energies. The factor $\sqrt2$ in \eqref{def_chi} is included to have $\hat \chi_q,\Phi_q$ and
 $\Psi_q$ all normalized to one. If the TI chain is homogeneous and critical, $h_i=\lambda_i=1$,  
 $\hat M$ is the same as $2 \hat H$ in \eqref{hamhop}. Therefore $\hat \chi_q$ has the form
 \eqref{eigen_chain}
 with $L \rightarrow 2L$, and due to \eqref{correl1} and \eqref{def_chi} the matrix $K_{m,n}$ and its
 transpose are submatrices of the $(2L\times 2L)$ correlation matrix $C$ of the hopping model
 \cite{Igloi/Juhasz08}
\begin{equation}
  K_{m,n}= -2C_{2m-1,2n} \,, \quad \quad K'_{m,n}= -2C_{2m,2n-1} \, ,
\label{correl2}
\end{equation}
where the minus sign appears because the ground state of the hopping model is half filled rather
than empty.

Finally, considering the subsystem of the first $2\ell$ sites in the hopping model and writing the
eigenvectors of $C_{i,j}$ in analogy to \eqref{def_chi} as
\begin{equation}
\chi_k=\left( \phi_k(1),\psi_k(1),\phi_k(2),\psi_k(2),\dots,\psi_k(\ell) \right)/\sqrt2\,,
\label{def_chi_sub}
\end{equation}
one sees that the equation $C \chi_k = \zeta_k \,\chi_k$, or
$(1-2C) \chi_k = \tanh(\varepsilon_k /2) \,\chi_k$, is equivalent to the system \eqref{diag2}
and can be used to obtain $\mathcal{H}$. The commuting matrix $T$, on the other hand, has the
structure of $\hat M$ in \eqref{tridiagonal3} with parameters
\begin{equation}
  h_i=t_{2i-1}\,,\quad \quad \lambda_i=t_{2i} \, ,
\label{comm_ti}
\end{equation}
where the $t_i$ are given by \eqref{coeff_final} or \eqref{coeff_chain_rescaled} with $L \rightarrow 2L$
and $\ell \rightarrow 2\ell$. Thus it describes a TI model on $\ell$ sites where couplings and fields
increase towards the boundary and where the Hamiltonian $\mathcal{T}$ commutes with $\mathcal{H}$.

The arguments used here are not restricted to the finite open chain, but hold also for rings or infinite
systems (with proper handling of the boundary conditions). Thus in all these cases commuting operators
follow from those of the half-filled hopping (XX) model.

\section*{References}

\providecommand{\newblock}{}


\begin{thebibliography}{100}
\expandafter\ifx\csname url\endcsname\relax
  \def\url#1{{\tt #1}}\fi
\expandafter\ifx\csname urlprefix\endcsname\relax\def\urlprefix{URL }\fi
\providecommand{\eprint}[2][]{\url{#2}}

\bibitem{CCD09}
Calabrese P, Cardy J and Doyon B 2009 Entanglement entropy in extended quantum systems
{\em J. Phys. A: Math. Theor. \/} {\bf 42} 500301

\bibitem{Laflorencie16}
Laflorencie N 2016 Quantum entanglement in condensed matter systems {\em Phys. Rep.\/} {\bf 643} 1

\bibitem{Li/Haldane08}
  Li H and Haldane F D M 2008 Entanglement spectrum as a generalization of entanglement entropy:
identification of topological order in non-abelian fractional quantum Hall effect states {\em Phys. Rev. Lett.\/} {\bf 101} 010504

  
\bibitem{Casini/Huerta/Myers11}
  Casini H, Huerta M and Myers R C 2011 Towards a derivation of holographic entanglement entropy
  {\em JHEP\/} {\bf 05} 036

\bibitem{Wong_etal13}
  Wong G, Klich I, Zayas L A P and Vaman D 2013 Entanglement temperature and entanglement entropy of excited states
  {\em JHEP\/} {\bf 12} 020 

\bibitem{Wen_etal16}
  Wen X, Ryu S and Ludwig A 2016 Evolution operators in conformal field theories and conformal mappings: entanglement Hamiltonian, the sine-square deformation, and others
  {\em Phys. Rev. B\/} {\bf 93} 235119
  
\bibitem{Cardy/Tonni16}
  Cardy J and Tonni E 2016 Entanglement Hamiltonians in two-dimensional conformal field theory
  {\em J. Stat. Mech.\/} P123103
  
\bibitem{Pretko17}
  Pretko M 2017 Nodal line entanglement entropy: Generalized Widom formula from entanglement Hamiltonians
  {\em Phys. Rev. B\/} {\bf 95} 235111

\bibitem{Arias_etal17_1}
  Arias R E, Blanco D D, Casini H and Huerta M 2017 Local temperatures and local terms in modular Hamiltonians
  {\em Phys. Rev. D\/} {\bf 95} 065005

\bibitem{Arias_etal17_2}
  Arias R E, Casini H, Huerta M and Pontello D 2017 Anisotropic Unruh temperatures
  {\em Phys. Rev. D\/} {\bf 96} 105019      

\bibitem{Klich/Vaman/Wong17_1}
  Klich I, Vaman D and Wong G 2017 Entanglement Hamiltonians for chiral fermions with zero modes
  {\em Phys. Rev. Lett.\/} {\bf 119} 120401

\bibitem{Klich/Vaman/Wong17_2}
  Klich I, Vaman D and Wong G 2018 Entanglement Hamiltonians and entropy in 1+1D chiral fermion systems
  {\em Phys. Rev. B\/} {\bf 98} 035134
    
\bibitem{Tonni/Laguna/Sierra17}
  Tonni E, Rodr\'{\i}guez-Laguna J and Sierra G 2018 Entanglement hamiltonian and entanglement contour in inhomogeneous 1D critical systems,
  {\em J. Stat. Mech.\/} 043105


\bibitem{Peschel/Eisler09}
  Peschel I and Eisler V 2009 Reduced density matrices and entanglement entropy in free lattice models
  {\em J. Phys. A: Math. Theor.\/} {\bf 42} 504003

\bibitem{Nienhuis/Campostrini/Calabrese09}
  Nienhuis B, Campostrini M and Calabrese P 2009 Entanglement, combinatorics and finite-size effects in spin-chains
  {\em J. Stat. Mech.\/} P02063
  
\bibitem{Kim_etal16}
  Kim P, Katsura H, Trivedi N and Han J H 2016 Entanglement and corner Hamiltonian spectra of integrable open spin chains
  {\em Phys. Rev. B\/} {\bf 94} 195110
  
\bibitem{Eisler/Peschel17}
  Eisler V and Peschel I 2017 Analytical results for the entanglement Hamiltonian of a free-fermion chain
  {\em J. Phys. A: Math. Theor.\/} {\bf 50} 284003

\bibitem{Dalmonte/Vermersch/Zoller17}
  Dalmonte M, Vermersch B and Zoller P  2018 Quantum Simulation and Spectroscopy of Entanglement Hamiltonians
  {\em Nat. Phys.\/} {\bf 14} 827

\bibitem{Parisen/Assaad18}
  Parisen Toldi F and Assaad F 2018 Entanglement Hamiltonian of interacting fermionic models,
  arXiv:1804.03163

  

\bibitem{Grünbaum81}
  Gr\"unbaum F~A 1981 Eigenvectors of a Toeplitz matrix: Discrete version of the prolate spheroidal wave functions
  {\em SIAM J. Alg. Disc. Meth.\/} {\bf 2} 136

\bibitem{Xu/Chamzas84}
  Xu W Y and Chamzas C 1984 On the periodic discrete prolate spheroidal sequences
  {\em SIAM J. Appl. Math.\/} {\bf 44} 1210 
  
\bibitem{Fagotti/Calabrese11}
  Fagotti M and Calabrese P 2011 Universal parity effects in the entanglement entropy of XX chains with open boundary conditions
  {\em J. Stat. Mech.\/} P01017

\bibitem{Igloi/Juhasz08}
  Igl\'oi F and Juh\'asz R  2008 Exact relationship between the entanglement entropies of XY and quantum Ising chains
  {\em EPL\/} {\bf 81} 57003
  
\bibitem{Peschel03}
  Peschel I 2003 Calculation of reduced density matrices from correlation functions
  {\em J. Phys. A: Math. Gen.\/} {\bf 36} L205
  
\bibitem{Eisler/Peschel13}
  Eisler V and Peschel I 2013 Free-fermion entanglement and spheroidal functions
  {\em J. Stat. Mech.\/} P04028
  
\bibitem{Slepian78}
  Slepian D 1978 Prolate spheroidal wave functions, Fourier analysis, and uncertainty V: the discrete case
  {\em Bell Syst. Techn. J.\/} {\bf 57} 1371
  
\bibitem{Peschel04}
  Peschel I 2004 On the reduced density matrix for a chain of free electrons {\em J. Stat. Mech.\/} P06004

\bibitem{Calabrese/Cardy04}
  Calabrese P and Cardy J 2004 Entanglement entropy and quantum field theory
  {\em J. Stat. Mech.\/} P06002
  
\bibitem{Dingle/Morgan67}
  Dingle R B and Morgan G J 1968 WKB methods for difference equations
  {\em Appl. Sci. Res.\/} {\bf 18} 221, 238
  
\bibitem{Grünbaum81a}
  Gr\"unbaum F~A 1981 Toeplitz matrices commuting with tridiagonal matrices {\em Lin. Alg. Appl.\/} {\bf 40} 25

  

\bibitem{Grady82}
  Grady M  1982 Infinite set of conserved charges in the Ising model
  {\em Phys. Rev. D\/} {\bf 25} 1103
 
\bibitem{Peschel/Schotte84}
  Peschel I and Schotte K D 1984 Time correlations in quantum spin chains and the X-ray absorption problem
  {\em Z. Physik B\/} {\bf 54} 305

\bibitem{Turban84}
  Turban L 1984 Exactly solvable spin-1/2 quantum chains with multispin interactions
  {\em Phys. Lett.\/} {\bf 104} 435

\bibitem{Igloi/Juhasz/Rieger00}
  Igl\'oi F, Juh\'asz R and Rieger H 2000 Random antiferromagnetic quantum spin chains: Exact results from scaling of rare regions
  {\em Phys. Rev. B\/} {\bf 61} 11552
                 
\bibitem{LSM61}
  Lieb E, Schultz T and Mattis D 1961  Two soluble models of an antiferromagnetic chain {\em Ann. Phys.\/} {\bf 16} 407      


\end{thebibliography}
\end{document}